\documentclass[12pt]{article}
\usepackage{amsmath}
\usepackage{amssymb}
\usepackage{babel}
\usepackage{latexsym}
\usepackage{times}
\usepackage{mathptm}
\begin{document}
\newcounter{saveeqn}
\newcommand{\alpheqn}{\setcounter{saveeqn}{\value{equation}}%
\setcounter{equation}{0}%
\renewcommand{\theequation}{\mbox{\arabic{saveeqn}-\alph{equation}}}}
\newcommand{\reseteqn}{\setcounter{equation}{\value{saveeqn}}%
\renewcommand{\theequation}{\arabic{equation}}}
\renewcommand{\thesection}{\Roman{section}}
\protect\newtheorem{principle}{Principle}[section]
\protect\newtheorem{theo}[principle]{Theorem}
\protect\newtheorem{prop}[principle]{Proposition}
\protect\newtheorem{lem}[principle]{Lemma}
\protect\newtheorem{co}[principle]{Corollary}
\protect\newtheorem{de}[principle]{Definition}
\newtheorem{ex}[principle]{Example}
\begin{titlepage} \enlargethispage{4cm}
\vspace*{-1.5cm}
\begin{center}
{\Large \bf Homogeneous decoherence functionals in standard \\ \vskip.06in}
{\Large \bf and history quantum mechanics}
\vskip.5in
{{\large Oliver Rudolph} $^a$}
\vskip.17in
{\normalsize Theoretical Physics Group}
\vskip.05in
{\normalsize Blackett Laboratory}
\vskip.05in
{\normalsize Imperial College of Science, Technology and Medicine}
\vskip.05in
{\normalsize South Kensington}
\vskip.05in
{\normalsize Prince Consort Road}
\vskip.05in
{\normalsize London SW7 2BZ, England}
\vskip.17in
{\normalsize and}
\vskip.17in
{{\large {J}.~{D}.~{Maitland Wright}} $^b$}
\vskip.17in
{\normalsize Analysis and Combinatorics Research Centre}
\vskip.05in
{\normalsize Mathematics Department}
\vskip.05in
{\normalsize University of Reading}
\vskip.05in
{\normalsize Reading RG6 6AX, England}
\vskip.45in
\end{center}
\normalsize
\begin{center}
{ABSTRACT}
\end{center}
\smallskip
\noindent  General history quantum theories are quantum theories
without a globally defined notion of time. Decoherence functionals
represent the states in the history approach and are defined
as certain bivariate complex-valued functionals on the space
of all histories. However, in practical situations -- for instance
in the history formulation of standard quantum mechanics -- there
often is a global time direction and the \emph{homogeneous} decoherence
functionals are specified by their values on the subspace of
homogeneous histories.

In this work we study the analytic properties of (i) the standard
decoherence functional in the history version of standard quantum
mechanics and (ii) \emph{homogeneous} decoherence functionals in general
history theories. We restrict ourselves to the situation where
the space of histories is given by the lattice of projections
on some Hilbert space $\mathcal{H}$.
Among other things we prove the non-existence
of a finitely valued extension for the standard decoherence functional
to the space of all histories, derive a representation for the
standard decoherence functional as an unbounded quadratic form
with a natural representation
on a Hilbert space and prove the existence of an Isham-Linden-Schreckenberg
(ILS) type representation
for the standard decoherence functional. \\
\\ \bigskip \noindent
\centerline{\vrule height0.25pt depth0.25pt width4cm \hfill}
\noindent
{\footnotesize $^a$ email: o.rudolph@ic.ac.uk \\
$^b$ email: J.D.M.Wright@reading.ac.uk}
\end{titlepage}
\newpage
\section{Introduction}
This paper is an investigation into certain aspects of the history
approach to quantum mechanics. The histories approach to quantum
theory is a promising new approach to quantum mechanics
\cite{Griffiths84} -
\cite{Craig97}
which has led to several interesting developments. Originally,
the \emph{consistent histories}
approach to quantum mechanics was introduced
by Griffiths \cite{Griffiths84}
as a tool for interpreting standard nonrelativistic
Hilbert space quantum mechanics. This so-called `consistent histories
interpretation' has been further developed and brought to its
present form by Omn\`{e}s \cite{Omnes94}. \\
However, in the present paper we are exclusively interested in
another aspect of the histories approach, namely that it is a
potential framework for a quantum theory where time plays a subsidiary
role. In a series of interesting papers Gell-Mann and Hartle
\cite{GellMannH90a} - \cite{GellMannH93}
have studied quantum cosmology and the path integral
formulation of relativistic quantum field theory in terms of
the concepts of the histories approach. They put forward for
the first time the idea of taking the concepts of the consistent
histories approach to quantum mechanics as independent fundamental
entities in their own right in a \emph{generalized quantum theory}.

Based on this idea, Isham has formulated in \cite{Isham94}
a natural algebraic
generalization of the consistent histories approach. With his
so-called \emph{general history quantum theories} he has broadened both
the scope and the mathematical framework of the consistent histories
approach to quantum mechanics. Standard quantum mechanics is
based on the idealized notions of observable and state at a single
time. Isham's general history theories provide a framework in
which these notions are replaced by \emph{temporal} analogues, histories
and decoherence functionals, respectively. Accordingly, in general
the histories are more general objects than simply time-sequences
of single-time events but are regarded as events intrinsically
spread out in time. \\
Moreover, Isham's general quantum histories provide a possible
framework for formulating a quantum theory without an external
globally defined notion of time. \\
In Isham's approach a general history quantum theory is formally
characterized by the space of histories on the one hand and by
the space of decoherence functionals on the other hand. The histories
are regarded as fundamental entities in their own right and are
identified with the general temporal properties of the quantum
system. Isham's approach has subsequently become the subject
of intense study \cite{IshamL94} - \cite{RudolphW97}.

In the history approach probabilities are assigned to complete
histories. However, at the basis of the histories approach is
the idea that any probability assignment to a history $h$ makes
sense only with respect to a so called \emph{consistent set} of histories
containing $h$. Dual to the notion of history is the notion of
decoherence functional. The decoherence functional determines
the consistent sets of histories in the theory and the probabilities
assigned to histories in the consistent sets. More specifically,
a decoherence functional $d$ maps every ordered pair of general
histories $h, k$ to a complex number denoted by $d(h,k)$. The number
$d(h,k)$ is interpreted in physical terms as a measure of the
mutual interference of the two histories $h$ and $k$. A consistent
set of histories consists of histories whose mutual interference
is sufficiently small, such that the diagonal value $d(h,h)$ can
be interpreted as the probability of the history $h$ in this consistent
set.

In the present work we study both the history version of standard
quantum mechanics and Isham's general history quantum theories.
In the latter case we restrict ourselves to the situation where
the space of histories is given by the space of projections on
some Hilbert space. In \cite{IshamLS94}
Isham, Linden and Schreckenberg studied
operator representations for decoherence functionals in the finite
dimensional case. For infinite dimensions, Wright \cite{Wright95}
obtained
the canonical representation of bounded decoherence functionals
by quadratic forms on von Neumann algebras. As a special case
of Wright's results in \cite{Wright95}, it is now known that a countably
additive decoherence functional, defined on all the projections
of an infinite dimensional separable Hilbert space, must be bounded.
In \cite{RudolphW97} we have further investigated operator representations
of bounded decoherence functionals in the infinite dimensional
case. However, in practical situations -- and in particular in
the history formulation of standard quantum mechanics -- decoherence
functionals are often specified by their values on the space
of so-called \emph{homogeneous histories},  which are simply time sequences
of single-time events (with respect to some \emph{a priori} given time
direction). The values of such a decoherence functional on the
space of all histories are in general unknown, and - moreover
- it is not at all clear \emph{a priori} whether such \emph{homogeneous
decoherence
functionals} can be extended to the space of all histories.

In this work we address the problem of whether such homogeneous
decoherence functionals can be extended unambiguously to the
set of all histories and study operator representations for homogeneous
decoherence functionals. For finite dimensional Hilbert spaces
we find in Section III that every bounded homogeneous decoherence
functional admits an Isham-Linden-Schreckenberg (ILS) representation
by some trace class operator and can be uniquely extended to
a bounded decoherence functional on the space of all histories.
In the infinite dimensional case we identify in Section IV those
homogeneous decoherence functionals admitting an ILS representation.
Section V is devoted to the study of the homogeneous decoherence
functional $d_\rho$ in the history version of standard quantum mechanics
corresponding to the initial state $\rho$. We shall show that the standard
homogeneous decoherence functional
$d_\rho$ \emph{cannot be represented} by a finitely
valued complex-valued (bounded or unbounded) decoherence functional
on the space of all histories whenever the single time Hilbert
space is infinite dimensional. Nevertheless we show that the
standard homogeneous decoherence functional admits a generalized
ILS-type representation by some bounded operator. Our result
shows that the standard decoherence functional - although bounded
on homogeneous histories - can only be extended to a function
on the space of all histories if values in the Riemann sphere
$\mathbb{C} \cup \{ \infty \}$
are permitted. Moreover, in V.2, we succeed in representing
the decoherence functional $d_\rho$ by an (in general) unbounded quadratic
form. This gives a very natural extension of Wright's representation
theory \cite{Wright95} for bounded decoherence functionals.
\section{Homogeneous decoherence functionals}
\subsection{Standard quantum mechanics}
In standard quantum mechanics single-time events at time $t$ are
represented by projection operators $h_t$ on the single-time Hilbert
space $\mathfrak{H}_s$
and the quantum mechanical state is given by some density
operator $\rho$ on $\mathfrak{H}_s$.
In the history formulation of nonrelativistic
quantum mechanics one considers \emph{homogeneous histories} which are
simply finite sequences $\{ h_t \}$ of single-time events parametrized
by the external time parameter $t$. Physically, one may think of
a homogeneous history as a sequence of quantum events, or --
in a measurement situation -- as a sequence of measurement results.

Let $h$ denote a finite (homogeneous) history, i.e., a finite sequence
$h_{t_1}, h_{t_2}, \cdots, h_{t_n}$ of projection operators $h_{t_j}$
on the single time Hilbert space $\mathfrak{H}_s$.
We call the number $n$ the \emph{order} of $h$. Then, by standard quantum
mechanics, the probability (symbolically denoted by $d^{hom}_\rho(h,h)$)
of the history $h$ in the quantum state $\rho$ is given by
\[ d_\rho^{hom}(h,h) = \mathrm{tr}_{\mathfrak{H}_s}(h_{t_n} h_{t_{n-1}}
\cdots h_{t_1} \rho h_{t_1} \cdots h_{t_{n-1}} h_{t_n}). \]
This expression for the probability of a homogeneous history
was first given by Wigner \cite{Wigner63} in the context of the orthodox
Copenhagen interpretation.

Since a quantum mechanical state $\rho$ may be identified with a positive
trace class operator on $\mathfrak{H}_s$ with trace one,
the expression for $d_\rho^{hom}(h,h)$ is well-defined,
even when some (or all) of the projection operators
$h_{t_j}$ have infinite dimensional range.

Accordingly, the standard homogeneous decoherence functional
$d_\rho^{hom}$ in standard quantum mechanics
(associated with the state $\rho$) is
defined for all pairs of finite homogeneous histories $h$ and $k$ as
\begin{equation}
d^{hom}_\rho(h,k) = \mathrm{tr}_{\mathfrak{H}_s}(h_{t_n} h_{t_{n-1}}
\cdots h_{t_1} \rho k_{t_1} \cdots k_{t_{n-1}} k_{t_n}).
\label{E1} \end{equation}
We can always assume without loss of generality that the order
of $h$ equals the order of $k$, whenever $h$ and $k$ are finite homogeneous
histories. We are working in the Heisenberg picture where the
time evolution is thrown into the projection operators
$h_{t_j}$ and --
to keep the notation as simple and as transparent as possible
-- the explicit time dependence of the projection operators
$h_{t_j}$ is suppressed.

The main idea in Isham's approach \cite{Isham94}
is to map homogeneous histories
injectively into the space of projection operators on some appropriate
($n$-fold) tensor product
$\mathcal{K}_{\: \: t_1, \cdots, t_n} :=
\otimes_{i=1}^n \mathfrak{H}_{t_i}$
where $\mathfrak{H}_{t_i}$ equals the single time Hilbert space
$\mathfrak{H}_s$ for every $i$. I.e., the history $\{ h_{t_j} \}$
is mapped to the projection operator
$h_{t_1} \otimes \cdots \otimes h_{t_n}$
on $\mathcal{K}_{\: \: t_1, \cdots, t_n}$.
We will normally follow Isham and identify a
homogeneous history with the corresponding projection operator
on the tensor product Hilbert space. Then, following Isham \cite{Isham94},
the standard decoherence functional $d_\rho$ is defined on the
pair
$\left(h_1 \otimes \cdots \otimes h_n, k_1 \otimes \cdots \otimes k_n
\right)$ as $d_\rho^{hom}(h_1, \cdots, h_n, k_1, \cdots, k_n)$. However,
it will be helpful, initially, to keep the distinction between
$d_\rho^{hom}$ and $d_\rho$ clear.

Assume that the spectral resolution of $\rho$ can be written as
$\rho = \sum_{i=1}^\infty \omega_i P_{\psi_i}$, where
$\{ \psi_i \}$ denotes an orthonormal basis in $\mathfrak{H}_s$, where
$P_{\psi_i}$ denotes the one dimensional
projection operator onto the subspace of
$\mathfrak{H}_s$ spanned by
$\psi_i$ for every $i$ and where $\sum_{i=1}^\infty \omega_i = 1$ and
$\omega_i \geq 0$ for all i. \\
Isham, Linden and Schreckenberg \cite{IshamLS94}
have shown that by repeatedly
inserting arbitrary `resolutions of the identity' into
$d^{hom}_\rho(h,k)$, the decoherence functional
$d^{hom}_\rho(h,k)$ can be written as
\begin{eqnarray*}
& d_\rho^{hom}(h,k) = & \! \! \!
\sum\limits_{i, j_1, \cdots, j_{2n+1} = 1}^\infty \left( \omega_i
\left\langle e^2_{j_2}, k_{t_1} e^1_{j_1} \right\rangle
\left\langle e^3_{j_3}, k_{t_2} e^2_{j_2} \right\rangle \cdots
\left\langle e^{n+1}_{j_{n+1}}, k_{t_n} e^n_{j_n} \right\rangle \times
\right. \\
& & \qquad \quad \hspace{2.5em} \times \left.
\left\langle e^{n+2}_{j_{n+2}}, h_{t_n} e^{n+1}_{j_{n+1}}
\right\rangle \cdots
\left\langle e^{2n+1}_{j_{2n+1}}, h_{t_1} e^{2n}_{j_{2n}}
\right\rangle
\left\langle e^1_{j_1}, P_{\psi_i} e^{2n+1}_{j_{2n+1}}
\right\rangle \right), \end{eqnarray*}
where the $\{ e^r_{j_r} \}$ are orthonormal bases in
$\mathfrak{H}_s$ for all $r$. Thus
\[ d_\rho^{hom}(h,k) = \sum_{i, j_2, \cdots, j_{2n}} \omega_i
\left\langle e^2_{j_2}, k_{t_1} \psi_i \right\rangle
\left\langle e^3_{j_3}, k_{t_2} e^2_{j_2} \right\rangle \cdots
\left\langle e^{n+1}_{j_{n+1}}, k_{t_n} e^n_{j_n} \right\rangle
\left\langle e^{n+2}_{j_{n+2}}, h_{t_n} e^{n+1}_{j_{n+1}}
\right\rangle \cdots
\left\langle \psi_i, h_{t_1} e^{2n}_{j_{2n}}
\right\rangle. \]
Hence we arrive at the representation for $d_\rho$
\begin{equation}
d_\rho(h,k) = \sum_{j_1, \cdots, j_{2n} = 1}^\infty \omega_{j_1}
\left\langle \left(h \otimes k \right) \left( \varepsilon_{j_1, \cdots,
j_{2n}} \right), \tilde{\varepsilon}_{j_1, \cdots, j_{2n}} \right\rangle
\label{E2} \end{equation}
for all homogeneous histories $h= h_{t_1} \otimes \cdots \otimes h_{t_n}$
and $k = k_{t_1} \otimes \cdots \otimes k_{t_n}$,
where the orthonormal bases $\{ \varepsilon_{j_1, \cdots,
j_{2n}} \}$ and $\{ \tilde{\varepsilon}_{j_1, \cdots, j_{2n}} \}$
of $\mathcal{K}_{\: \: t_1, \cdots, t_n} \otimes \mathcal{K}_{\: \: t_1,
\cdots, t_n}$
are given by \alpheqn \begin{eqnarray}
\varepsilon_{j_1, \cdots, j_{2n}} & := & \psi_{j_1} \otimes
e^{2n}_{j_{2n}} \otimes e^{2n-1}_{j_{2n-1}} \otimes \cdots \otimes
e^{n+2}_{j_{n+2}} \otimes e^2_{j_2} \otimes e^3_{j_3} \otimes
\cdots \otimes e^{n+1}_{j_{n+1}} \label{E2a} \\
\tilde{\varepsilon}_{j_1, \cdots, j_{2n}} & := &
e^{2n}_{j_{2n}} \otimes e^{2n-1}_{j_{2n-1}} \otimes
\cdots \otimes e^{n+1}_{j_{n+1}} \otimes \psi_{j_1} \otimes
e^2_{j_2} \otimes e^3_{j_3}
\otimes \cdots \otimes e^n_{j_n} \label{E2b}
\end{eqnarray} \reseteqn
The expression (\ref{E2}) is well-defined and finite for all pairs
of homogeneous histories $h= h_{t_1} \otimes \cdots \otimes h_{t_n}$
and $k = k_{t_1} \otimes \cdots \otimes k_{t_n}$.

Following Isham \cite{Isham94}
in the history reformulation of quantum mechanics
the set of all histories now has to be identified with the set
of projections $\mathcal{P}\left(\mathcal{K}_{\: \: t_1, \cdots, t_n}
\right)$ of projection operators on
$\mathcal{K}_{\: \: t_1, \cdots, t_n}$. [Strictly speaking
the set of all histories has to be identified with the direct
limit of the directed system $\left\{ \mathcal{P} \left(
\mathcal{K}_{\: \: t_1, \cdots, t_n} \right) : \{ t_1, \cdots, t_n \}
\subset \mathbb{R} \right\}$, see \cite{Pulmannova95,Rudolph96}.]
Those histories in $\mathcal{P} \left(\mathcal{K}_{\: \: t_1, \cdots, t_n}
\right)$ which are not homogeneous are called
\emph{inhomogeneous histories}.
\subsection{General history quantum theories}
Now we switch to general (abstract) history quantum theories
over some Hilbert space $\mathcal{H}$. Such a theory is fully characterized
by the space of histories and the space of decoherence functionals.
In general history quantum theories over some Hilbert space $\mathcal{H}$
the space of all histories (or more precisely propositions about
histories) is -- by definition -- given by $\mathcal{P}(\mathcal{H})$,
see \cite{Isham94} - \cite{IshamL95}.
Notice that the history Hilbert space $\mathcal{H}$ must not be confused
with the single time Hilbert space $\mathfrak{H}_s$
in standard quantum mechanics.

A \emph{generalized decoherence functional} for $\mathcal{H}$
is a function $d$, defined
on all ordered pairs of projections in $\mathcal{P}(\mathcal{H})$,
with values in the
Riemann sphere $\mathbb{C} \cup \{ \infty \}$ such that:
\begin{itemize}
\item[(i)] $d(p,q) = d(q,p)^*$ for each $p$ and each $q$ in
$\mathcal{P}(\mathcal{H})$. (Hermitianness)
\item[(ii)] $d(p,p) \geq 0$ for each $p$. (Positivity)
\item[(iii)] $d(1,1) = 1$. (Normalization)
\item[(iv)] $d(p_1 + p_2 , q) = d(p_1 ,q)
+ d(p_2 ,q)$  for each $q$ whenever $p_1$ and $p_2$
are perpendicular and all quantities
and terms are finite. (Ortho-additivity).
\end{itemize}
Moreover, we say that a decoherence functional $d$ is completely
additive if \begin{itemize}
\item[(iv')] whenever $\{ p_i \}_{i \in I}$
is an infinite collection of pairwise orthogonal projections, \[
d \left( \sum_{i \in I} p_i, q \right) = \sum_{i \in I} d(p_i, q),
\] for all $q \in \mathcal{P}(\mathcal{H})$
such that the left hand side is finite and all terms
in the summation on the right hand side are finite.
The infinite series is required to converge absolutely. [This is
automatic when $I$ is countable since the series is rearrangement
invariant. When $I$ is uncountable, all but countably many of the
terms of the series are zero.]
\end{itemize}
For brevity we shall write `decoherence functional' for `generalized
decoherence functional' except where this could cause confusion.
In the previous literature it has always been assumed that decoherence
functionals are finitely valued. In the present work, however,
we drop the requirement that decoherence functionals are finitely
valued. Our motivation for doing so will become clear below,
see Section V.

In the present paper we restrict ourselves to the situation where
the history Hilbert space $\mathcal{H}$ can be written as a finite tensor
product of a family of Hilbert spaces. Specifically we are aiming
at formalising those situations where there is given \emph{a priori} an
external (possibly discretised or coarse grained) time direction
in the theory -- as for instance in the history formulation of
standard quantum mechanics. In this case the notions of homogeneous
history and of homogeneous decoherence functional make sense.

The history reformulation of standard quantum mechanics discussed
above motivates also the following general definitions for general
history quantum theories. \\ \\
\textbf{Definition} \emph{Let
$\mathcal{B}(\mathcal{H})$ be the (von Neumann) algebra of all bounded
operators on a Hilbert space $\mathcal{H}$ and
let $\mathcal{P}(\mathcal{H})$ be the lattice of
projections in $\mathcal{B}(\mathcal{H})$.
Assume that there is a finite family of Hilbert spaces $\{ \mathfrak{H}_i
\}_{i=1}^n$ such that $\mathcal{H}$
can be written as the tensor product of the Hilbert
spaces $\mathfrak{H}_i$, i.e., \[ \mathcal{H} = \otimes_{i=1}^n
\mathfrak{H}_i. \] Then we say that the pair $( \mathcal{H}, \{
\mathfrak{H}_i \} )$ is a \textbf{homogeneous history
Hilbert space of order} $n$.
In this situation we will -- abusing
language -- also briefly say that $\mathcal{H}$
is a homogeneous history Hilbert
space. When we do so we will always tacitly assume that a family
of Hilbert spaces $\{ \mathfrak{H}_i \}$ has been chosen such that
$(\mathcal{H}, \{ \mathfrak{H}_i \})$
is a homogeneous history Hilbert space. \\
A \textbf{homogeneous projection} $p$ on a
homogeneous history Hilbert space
$\mathcal{H}$ is then a projection of the form $p =
p_1 \otimes \cdots \otimes p_n$, where $p_i$ is a projection
on $\mathfrak{H}_i$ for all $i$ respectively.
A \textbf{homogeneous decoherence functional}
for $\mathcal{H}$ is a complex valued function $d^{hom}$,
defined on all pairs of $n$th order history projections,
i.e., $d^{hom}$ is a function
$d^{hom} : \mathcal{P}(\mathfrak{H}_1) \times \cdots \times
\mathcal{P}(\mathfrak{H}_n) \times \mathcal{P}(\mathfrak{H}_1)
\times \cdots \times
\mathcal{P}(\mathfrak{H}_n) \to \mathbb{C}, $ such that:
\begin{itemize}
\item[(i)] $d^{hom}(p_1, \cdots, p_n, q_1, \cdots, q_n) = d^{hom}(q_1,
\cdots, q_n, p_1, \cdots, p_n)^*$ for all $p_i$ and $q_i$ in
$\mathcal{P}(\mathfrak{H}_i)$. \emph{(Hermitianness)}
\item[(ii)] $d^{hom}(p_1, \cdots, p_n, p_1, \cdots, p_n) \geq 0$
for all $(p_1, \cdots, p_n)$. \emph{(Positivity)}
\item[(iii)] $d^{hom}(1_1, \cdots, 1_n, 1_1, \cdots, 1_n) = 1$.
\emph{(Normalization)}
\item[(iv)] $d^{hom}$ is
orthoadditive in each of its $2n$ arguments. \emph{(Ortho-additivity)}.
\\ \end{itemize}}
Notice that we only consider finitely valued homogeneous decoherence
functionals.

Clearly, in the physical applications in standard quantum mechanics
the Hilbert spaces
$\mathfrak{H}_i$  are all interpreted as the single time Hilbert
space $\mathfrak{H}_s$
indexed by a discrete time parameter, i.e., all the
$\mathfrak{H}_i$ are isomorphic to $\mathfrak{H}_s$
and can be obtained from the single-time
Hilbert space $\mathfrak{H}_s$
at some fiducial time by the application of
a suitable unitary time translation operator (recall that we
are working in the Heisenberg picture). For every initial state
$\rho$ the homogeneous decoherence functional
$d_\rho^{hom}$ associated with $\rho$ is
a homogeneous decoherence functional in the above sense.
From our discussion above it is clear that the homogeneous decoherence
functional in standard quantum mechanics is bounded. It is thus
of some interest to study the problem whether bounded homogeneous
decoherence functionals in general history quantum theories can
be unambiguously extended to the space of all histories.

In the sequel we shall need the following theorem.
\begin{theo} \label{T21}
Let $\mathcal{H} = \otimes_{i=1}^n \mathfrak{H}_i$ be
a history Hilbert space where all
$\mathfrak{H}_i$ are
of dimension greater than two. Let $d^{hom}$ be a bounded homogeneous
decoherence functional for $\mathcal{H}$.
Then there exists a unique bounded multilinear functional $\mathcal{D} :
\mathcal{B}(\mathfrak{H}_1) \times \cdots \times
\mathcal{B}(\mathfrak{H}_n) \times \mathcal{B}(\mathfrak{H}_1)
\times \cdots \times
\mathcal{B}(\mathfrak{H}_n) \to \mathbb{C}$
extending $d^{hom}$. \end{theo}
This theorem is a special case of the multi-form generalized
Gleason theorem proved in \cite{RudolphW98}.

In the next section we will briefly consider the situation where
all $\mathfrak{H}_i$ (and thus also $\mathcal{H} = \otimes_{i=1}^n
\mathfrak{H}_i$) are finite dimensional Hilbert spaces
whereas in the Sections IV and V we consider the situation where
the $\mathfrak{H}_i$ are general infinite dimensional Hilbert spaces.
\section{The
finite dimensional case: \\ a generalized Isham-Linden-Schreckenberg
Theorem}
In this section we briefly consider the question whether in history
quantum theories over finite dimensional Hilbert spaces every
bounded homogeneous decoherence functional can be extended to
a finitely valued decoherence functional on the space of all
histories. This question can indeed be answered in the affirmative.
\begin{theo}
Let $\mathcal{H}$ be a finite dimensional homogeneous history
Hilbert space and $\mathcal{H} = \otimes_{i=1}^n \mathfrak{H}_i$
its representation as a finite tensor
product of (finite dimensional) Hilbert spaces all of which have
dimension greater than two. Then there is a one-to-one correspondence
between bounded homogeneous decoherence functionals $d^{hom}$ for
$\mathcal{H}$ and trace class operators $\mathfrak{X}$
on $\mathcal{H} \otimes \mathcal{H}$ according to the rule
\[ d^{hom}(p_1, \cdots, p_n, q_1, \cdots, q_n) =
\mathrm{tr}_{\mathcal{H} \otimes \mathcal{H}} \left( (p_1 \otimes
\cdots \otimes p_n \otimes q_1 \cdots \otimes q_n) \mathfrak{X}
\right) \] for all projections $p_j, q_j \in \mathcal{P}(\mathfrak{H}_j)$
with the restriction that \begin{itemize}
\item[(i)] $\mathrm{tr}_{\mathcal{H} \otimes \mathcal{H}} \left(
(p_1 \otimes
\cdots \otimes p_n \otimes q_1 \cdots \otimes q_n) \mathfrak{X}
\right) = \mathrm{tr}_{\mathcal{H} \otimes \mathcal{H}} \left(
(q_1 \otimes
\cdots \otimes q_n \otimes p_1 \cdots \otimes p_n) \mathfrak{X}^*
\right);$
\item[(ii)] $\mathrm{tr}_{\mathcal{H} \otimes \mathcal{H}} \left(
(p_1 \otimes
\cdots \otimes p_n \otimes p_1 \cdots \otimes p_n) \mathfrak{X}
\right) \geq 0$;
\item[(iii)] $\mathrm{tr}_{\mathcal{H} \otimes \mathcal{H}} \left(
\mathfrak{X} \right) =1$.
\end{itemize} \label{T31} \end{theo}
\textbf{Proof}:
This theorem can be proved directly by iterating the argument
given by Isham, Linden and Schreckenberg in their proof of the
case $n=1$ (the ILS-Theorem) \cite{IshamLS94}.
The details are left to the
reader. In Section IV we will derive Theorem \ref{T31} as a by-product
of Theorem \ref{T41}. \hfill $\Box$ \\ \\
\emph{Remark}: From Theorem \ref{T31} and from the ILS-Theorem
\cite{IshamLS94} it follows
that in the finite dimensional case the notions of bounded homogeneous
decoherence functional and bounded decoherence functional can
be identified with each other.
\section{The ILS-Theorem for homogeneous decoherence functionals in
infinite dimensions}
Let $\mathcal{H}$ be a Hilbert space and let
$\mathcal{K}(\mathcal{H})$ be the ideal of compact
operators in $\mathcal{B}(\mathcal{H})$.
Then $\mathcal{K}(\mathcal{H}) = \mathcal{B}(\mathcal{H})$ if,
and only if, $\mathcal{H}$ is finite
dimensional.

We shall need some basic facts on tensor products of operator
algebras. For a particularly elegant account, from first principles,
of tensor products of $C^*$-algebras see Wegge-Olsen \cite{WeggeOlsen92}
and, for
a more advanced treatment, see Kadison and Ringrose \cite{KadisonR8386}.
Let us recall that if $\mathfrak{H}_1, \cdots, \mathfrak{H}_n$
are Hilbert spaces, the algebraic tensor product
$\mathfrak{H}_1 \otimes_{alg} \cdots \otimes_{alg} \mathfrak{H}_n$
can be equipped with an inner product such that
$\left\langle \varphi_1 \otimes \cdots \otimes \varphi_n, \psi_1
\otimes \cdots \otimes \psi_n \right\rangle = \left\langle \varphi_1, \psi_1
\right\rangle \cdots \left\langle \varphi_n, \psi_n
\right\rangle$. The completion of
$\mathfrak{H}_1 \otimes_{alg} \cdots \otimes_{alg} \mathfrak{H}_n$
with respect to this inner product is the Hilbert
space tensor product
$\mathfrak{H}_1 \otimes \cdots \otimes \mathfrak{H}_n$.
When $\left\{ x_j \right\}_{j=1}^n$ is a family of bounded operators
on $\mathfrak{H}_j$ respectively, then there is a unique operator in
$\mathcal{B}(\mathfrak{H}_1 \otimes \cdots \otimes
\mathfrak{H}_n)$, denoted by
$x_1 \otimes \cdots \otimes x_n$, such that $(x_1 \otimes \cdots \otimes
x_n) (\varphi_1 \otimes \cdots \otimes \varphi_n) = x_1(\varphi_1)
\otimes \cdots \otimes x_n(\varphi_n)$.

Let $\{ \mathcal{A}_j \}_{j=1}^n$ be a
family of $C^*$-algebras of operators acting on
$\mathfrak{H}_j$ respectively.
Then the \emph{algebraic tensor product}
$\mathcal{A}_1 \otimes_{alg} \cdots \otimes_{alg} \mathcal{A}_n$
can be identified with the
$^*$-algebra, acting on
$\mathfrak{H}_1 \otimes \cdots \otimes \mathfrak{H}_n$,
which consists of all finite sums of operators
of the form $x_1 \otimes \cdots \otimes x_n$, with
$x_j \in \mathcal{A}_j$, for all $j$. The norm closure of
\mbox{$\mathcal{A}_1 \otimes_{alg} \cdots \otimes_{alg}
\mathcal{A}_n$}
is the $C^*$-\emph{tensor product of}
$\{ \mathcal{A}_j \}_{j=1}^n$ and is denoted by
$\mathcal{A}_1 \otimes \cdots \otimes \mathcal{A}_n$.
(This is also called
the spatial $C^*$-tensor product to distinguish it from other possible
$C^*$-tensor products; see \cite{WeggeOlsen92}.)

The algebraic tensor product
$\mathcal{K}(\mathfrak{H}_1) \otimes_{alg} \cdots \otimes_{alg}
\mathcal{K}(\mathfrak{H}_n)$ embeds naturally into
$\mathcal{B}(\mathfrak{H}_1 \otimes \cdots \otimes
\mathfrak{H}_n)$ (see, e.g., Kadison and Ringrose
\cite{KadisonR8386}, Chapter 11.2). This embedding of
$\mathcal{K}(\mathfrak{H}_1) \otimes_{alg} \cdots \otimes_{alg}
\mathcal{K}(\mathfrak{H}_n)$ in
$\mathcal{B}(\mathfrak{H}_1 \otimes \cdots \otimes
\mathfrak{H}_n)$ induces a (unique)
pre-$C^*$-norm on
$\mathcal{K}(\mathfrak{H}_1) \otimes_{alg} \cdots \otimes_{alg}
\mathcal{K}(\mathfrak{H}_n)$. The (spatial) $C^*$-tensor product
$\mathcal{K}(\mathfrak{H}_1) \otimes \cdots \otimes
\mathcal{K}(\mathfrak{H}_n)$ is the closure of
$\mathcal{K}(\mathfrak{H}_1) \otimes_{alg} \cdots \otimes_{alg}
\mathcal{K}(\mathfrak{H}_n)$ in
$\mathcal{B}(\mathfrak{H}_1 \otimes \cdots \otimes
\mathfrak{H}_n)$ with respect to this pre-$C^*$-norm
and can be identified with
$\mathcal{K}(\mathfrak{H}_1 \otimes \cdots \otimes
\mathfrak{H}_n)$.

Now let $\mathcal{H}$ be a homogeneous history Hilbert
space of order $n>0$ and let $\mathcal{H} = \otimes_{i=1}^n \mathfrak{H}_n$
be its given representation as a tensor product,
where all $\mathfrak{H}_i$ are of dimension greater than two.
Let $d^{hom}$ be a bounded
homogeneous decoherence functional for $\mathcal{H}$.
Then, by Theorem \ref{T21},
there exists a (unique) bounded $2n$-linear functional
$B : \mathcal{B}(\mathfrak{H}_1) \times \cdots \times
\mathcal{B}(\mathfrak{H}_n) \times
\mathcal{B}(\mathfrak{H}_1) \times \cdots \times
\mathcal{B}(\mathfrak{H}_n) \to \mathbb{C}$
such that $d^{hom}(p_1, \cdots, p_n, q_1, \cdots, q_n) = B(p_1,
\cdots, p_n, q_1, \cdots, q_n)$ for all $p_i, q_i \in
\mathcal{P}(\mathcal{H})$.

Denote by $B_K$ the restriction of $B$ to
$\mathcal{K}(\mathfrak{H}_1) \times \cdots \times
\mathcal{K}(\mathfrak{H}_n) \times
\mathcal{K}(\mathfrak{H}_1) \times \cdots \times
\mathcal{K}(\mathfrak{H}_n)$. Then, by the fundamental
property of the algebraic tensor product, there is a unique linear
functional \mbox{$\beta :
\mathcal{K}(\mathfrak{H}_1) \otimes_{alg} \cdots \otimes_{alg}
\mathcal{K}(\mathfrak{H}_n) \otimes_{alg}
\mathcal{K}(\mathfrak{H}_1) \otimes_{alg} \cdots \otimes_{alg}
\mathcal{K}(\mathfrak{H}_n) \to \mathbb{C}$} such that
\[ \beta(x_1 \otimes \cdots \otimes x_n \otimes y_1 \otimes \cdots
\otimes y_n) = B_K(x_1, \cdots, x_n, y_1, \cdots, y_n) =
B(x_1, \cdots, x_n, y_1, \cdots, y_n) \] for all
$x_i, y_i \in \mathcal{K}(\mathfrak{H}_i)$. In particular
$d^{hom}(p_1, \cdots, p_n, q_1, \cdots, q_n) =
\beta(x_1 \otimes \cdots \otimes x_n \otimes y_1 \otimes \cdots
\otimes y_n)$ for all projections $p_i, q_i \in
\mathcal{K}(\mathfrak{H}_i)$. \\ \\
\noindent \textbf{Definition} \emph{A
homogeneous decoherence functional $d^{hom}$ for a history
Hilbert space $\mathcal{H} = \otimes_{i=1}^n \mathfrak{H}_i$
is said to be \textbf{tensor bounded} if $d^{hom}$ is bounded
and the associated functional $\beta$ is \linebreak[3] bounded on
$\mathcal{K}(\mathfrak{H}_1) \otimes_{alg}
\mathcal{K}(\mathfrak{H}_2) \otimes_{alg} \cdots \otimes_{alg}
\mathcal{K}(\mathfrak{H}_n) \otimes_{alg}
\mathcal{K}(\mathfrak{H}_1) \otimes_{alg}
\mathcal{K}(\mathfrak{H}_2) \otimes_{alg} \cdots \otimes_{alg}
\mathcal{K}(\mathfrak{H}_n)$, when \linebreak[4]
\mbox{$\mathcal{K}(\mathfrak{H}_1) \otimes \cdots \otimes
\mathcal{K}(\mathfrak{H}_n) \otimes
\mathcal{K}(\mathfrak{H}_1) \otimes \cdots \otimes
\mathcal{K}(\mathfrak{H}_n)$} is equipped with its canonical
$C^*$-norm.}

\begin{theo}
Let $\mathcal{H}$ be a history Hilbert space with tensor product
representation $\mathcal{H} = \otimes_{i=1}^n \mathfrak{H}_i$
where all $\mathfrak{H}_i$ are of dimension greater than
two. Let $d^{hom}$ be a bounded homogeneous decoherence functional
for $\mathcal{H}$. Then $d^{hom}$ is tensor bounded if,
and only if, there exists a trace class operator $T$ on
$\mathfrak{H}_1 \otimes \cdots \otimes
\mathfrak{H}_n \otimes \mathfrak{H}_1 \otimes \cdots \otimes
\mathfrak{H}_n$ such that \[ d^{hom}(p_1, \cdots, p_n, q_1,
\cdots, q_n) = \mathrm{tr}_{\mathcal{H} \otimes \mathcal{H}}\left(
(p_1 \otimes \cdots \otimes p_n \otimes q_1 \otimes \cdots \otimes
q_n) T \right) \] for all projections
$p_i, q_i \in \mathcal{K}(\mathfrak{H}_i)$. \label{T41} \end{theo}
\textbf{Proof}: The proof is analogous to the proof of Theorem 3.2 in
\cite{RudolphW97} and omitted. \hfill $\Box$
\begin{co} \label{C42}
Let $\mathcal{H}$ be a history Hilbert space with tensor product
representation $\mathcal{H} = \otimes_{i=1}^n \mathfrak{H}_i$
where all $\mathfrak{H}_i$ are of dimension greater than
two. Let $d^{hom}$ be a completely additive bounded homogeneous
decoherence functional for $\mathcal{H}$. Then
$d^{hom}$ is tensor bounded if, and
only if, there exists a trace
class operator $T$ on
$\mathfrak{H}_1 \otimes \cdots \otimes
\mathfrak{H}_n \otimes \mathfrak{H}_1 \otimes \cdots \otimes
\mathfrak{H}_n$ such that
\begin{equation} \label{E3}
d^{hom}(p_1, \cdots, p_n, q_1,
\cdots, q_n) = \mathrm{tr}_{\mathcal{H} \otimes \mathcal{H}}\left(
(p_1 \otimes \cdots \otimes p_n \otimes q_1 \otimes \cdots \otimes
q_n) T \right) \end{equation} for all projections
$p_i, q_i \in \mathcal{P}(\mathfrak{H}_i)$. \end{co}
\textbf{Proof}: When there exists a trace class operator $T$ such that
(\ref{E3}) holds, then, by Theorem \ref{T41}, $d^{hom}$
is tensor bounded. Conversely,
when $d^{hom}$ is tensor bounded, the existence of $T$ such that
(\ref{E3})
holds for all projections of finite rank is guaranteed by Theorem
\ref{T41}. By appealing to the complete additivity of $d^{hom}$ and the
ultraweak continuity of the map $z \mapsto \mathrm{tr}(zT)$
it is straightforward
to establish (\ref{E3}) for arbitrary projections. \hfill $\Box$
\begin{co}
\label{C43}
Let $\mathcal{H}$ be a history Hilbert space with tensor product
representation $\mathcal{H} = \otimes_{i=1}^n \mathfrak{H}_i$
where all $\mathfrak{H}_i$ are of dimension greater than
two. There is a one-to-one correspondence between completely
additive, tensor bounded homogeneous decoherence functionals
for $\mathcal{H}$ and trace class operators $T$ on
$\mathfrak{H}_1 \otimes \cdots \otimes
\mathfrak{H}_n \otimes \mathfrak{H}_1 \otimes \cdots \otimes
\mathfrak{H}_n$ such that, for $p_i, q_i \in \mathcal{P}(\mathfrak{H}_i)$,
\begin{itemize}
\item $\mathrm{tr}_{\mathcal{H} \otimes \mathcal{H}}\left(
(p_1 \otimes \cdots \otimes p_n \otimes q_1 \otimes \cdots \otimes
q_n) T \right) = \mathrm{tr}_{\mathcal{H} \otimes \mathcal{H}}\left(
(q_1 \otimes \cdots \otimes q_n \otimes p_1 \otimes \cdots \otimes
p_n) T^* \right)$;
\item $\mathrm{tr}_{\mathcal{H} \otimes \mathcal{H}}\left(
(p_1 \otimes \cdots \otimes p_n \otimes p_1 \otimes \cdots \otimes
p_n) T \right) \geq 0$;
\item $\mathrm{tr}_{\mathcal{H} \otimes \mathcal{H}}\left(T \right)
=1$. \end{itemize}
\end{co}
\textbf{Proof}: Straightforward. \hfill $\Box$ \\ \\
\emph{Remark}: Theorem \ref{T31}
(the generalized Isham-Linden-Schreckenberg
Theorem) follows immediately since, when the history Hilbert
space $\mathcal{H}$ is finite dimensional, then
$\mathcal{K}(\mathfrak{H}_1) \otimes_{alg} \cdots \otimes_{alg}
\mathcal{K}(\mathfrak{H}_n) =
\mathcal{K}(\mathfrak{H}_1) \otimes \cdots \otimes
\mathcal{K}(\mathfrak{H}_n) =
\mathcal{B}(\mathfrak{H}_1 \otimes \cdots \otimes
\mathfrak{H}_n)$  which is finite dimensional
and every linear functional on a finite dimensional normed space
is bounded. \\

Let $d$ be a bounded decoherence functional for $(\mathcal{H}, \{
\mathfrak{H}_i \})$ where all $\mathfrak{H}_i$
have dimension greater than two. Let us call $d$
\emph{Isham-Linden-Schreckenberg-representable}
(or, more shortly, \emph{ILS-representable}) if there exists a trace
class operator $T$ in $\mathcal{B}(\mathcal{H} \otimes \mathcal{H}) =
\mathcal{B}(\mathfrak{H}_1 \otimes \cdots \otimes
\mathfrak{H}_n \otimes
\mathfrak{H}_1 \otimes \cdots \otimes
\mathfrak{H}_n)$ such that
\[ d(p,q) = \mathrm{tr}_{\mathcal{H} \otimes
\mathcal{H}}((p \otimes q)T) \]
for all projections $p$ and $q$ in $\mathcal{B}(\mathcal{H})$.
It follows
from the results given above that a completely additive homogeneous
decoherence functional $d^{hom}$ on a history Hilbert space $\mathcal{H}$
can be represented by an ILS-representable bounded decoherence functional
$d$ on $\mathcal{H}$ if, and only if, $d^{hom}$ is tensor bounded.
\section{The decoherence functional in standard quantum mechanics}
\subsection{Non-existence of a finitely valued extension of the standard
decoherence functional} \label{S51}
The definition of the homogeneous decoherence functional
$d^{hom}_\rho$ associated
with the initial state $\rho$ in the history reformulation of standard
quantum mechanics has been already given in Section II.1 above.
This function is of particular interest since the axioms characterizing
general history quantum theories are abstracted from the structure
of the history reformulation of standard quantum mechanics, and
features which fail to be true in the history reformulation of
standard quantum mechanics are unlikely to hold for more general
physical history quantum theories.

We recall that in standard quantum mechanics the homogeneous
decoherence functional $d^{hom}_\rho$
associated with the initial state $\rho$ is
defined on pairs of homogeneous histories $h$ and $k$ by Equation
(\ref{E1}) as
$d^{hom}_\rho(h,k) = \mathrm{tr}_{\mathfrak{H}_s}(h_{t_n} h_{t_{n-1}}
\cdots h_{t_1} \rho k_{t_1} \cdots k_{t_{n-1}} k_{t_n})$.

We have discussed in Section II.1 that in Isham's history formulation
of standard quantum mechanics the general histories are identified
with the projection operators on some tensor product of the single
time Hilbert space by itself. As explained in Section II.1 in
the tensor product formalism Isham et al.~\cite{IshamLS94}
obtained the representation in Equation (\ref{E2}) for
$d_\rho$
\[ d_\rho(h,k) = \sum_{j_1, \cdots, j_{2n} = 1}^\infty \omega_{j_1}
\left\langle \left(h \otimes k \right) \left( \varepsilon_{j_1, \cdots,
j_{2n}} \right), \tilde{\varepsilon}_{j_1, \cdots, j_{2n}}
\right\rangle \]
for all tensored homogeneous histories $h = h_{t_1} \otimes \cdots h_{t_n}$
and $k = k_{t_1} \otimes \cdots h_{t_n}$, where $\rho = \sum_{i=1}^\infty
\omega_i P_{\psi_i}$ denotes the spectral resolution of $\rho$ and where
the $\varepsilon_{j_1, \cdots, j_{2n}}$ and $\tilde{\varepsilon}_{j_1,
\cdots, j_{2n}}$ are defined in Equations (\ref{E2a}) and
(\ref{E2b}) respectively.

The question which will be addressed in this section is whether
this expression can be extended to the space of all histories.
When the single-time Hilbert space is finite dimensional, then
it follows from our result in Section III that $d_\rho$ can indeed be
extended to the space of all histories and its extension is also
ILS-representable.

It is natural to try to define $d_\rho(p,q)$ for arbitrary
projections $p,q$ by
\begin{equation}
\label{E4}
d_\rho(p,q) = \sum_{j_1, \cdots, j_{2n} = 1}^\infty \omega_{j_1}
\left\langle \left(p \otimes q \right) \left(
\varepsilon_{j_1, \cdots,
j_{2n}} \right), \tilde{\varepsilon}_{j_1, \cdots, j_{2n}}
\right\rangle. \end{equation}
\begin{prop}
\label{P51} When the single time Hilbert space is infinite
dimensional, the expression (\ref{E4}) does not define a finitely valued
functional on the space of all histories.
\end{prop}
\textbf{Proof}: Let $\mathfrak{H}_s$
denote the single-time Hilbert space of the quantum
system in question. We assume that
$\mathfrak{H}_s$ is a separable infinite dimensional
Hilbert space. We consider two time histories, i.e., the case
$n = 2$. Let $e$ be a fixed unit vector in $\mathfrak{H}_s$.
Let $(e_j)(j = 1,2...)$
be an orthonormal basis in $\mathfrak{H}_s$
with $e = e_1$. Let $P$ and $Q$ be projections
on $\mathfrak{H}_s \otimes \mathfrak{H}_s$.
Whenever the summation converges we define
\begin{eqnarray*}
d_e(P,Q) & := & \sum_{j(1)=1}^\infty \sum_{j(2)=1}^\infty
\sum_{j(3)=1}^\infty \left\langle (P \otimes Q)
(e \otimes e_{j(1)} \otimes e_{j(2)} \otimes e_{j(3)}), e_{j(1)}
\otimes e_{j(3)} \otimes e \otimes e_{j(2)} \right\rangle \\
& = & \sum_{j(1)=1}^\infty \sum_{j(2)=1}^\infty
\sum_{j(3)=1}^\infty \left\langle P (e \otimes e_{j(1)}), e_{j(1)}
\otimes e_{j(3)} \right\rangle \left\langle Q (e_{j(2)} \otimes e_{j(3)}),
e \otimes e_{j(2)} \right\rangle. \end{eqnarray*}
This expression coincides with the above formula defining the
standard decoherence functional for $n=2$ and for a pure quantum
state $\rho = P_e$, where $P_e$ denotes the projection
operator onto the subspace of $\mathfrak{H}_s$
spanned by $e$. If we put $P = I$, then all the terms in the
summation vanish except where $j(1) =j(3) = 1$. Thus
\[ d_e(I,Q) = \sum_{j(2)=1}^\infty
\left\langle Q (e_{j(2)} \otimes e), e \otimes e_{j(2)}
\right\rangle. \]
Let $D_e(S) := \sum_{j(2)=1}^\infty
\left\langle S (e_{j(2)} \otimes e), e \otimes e_{j(2)}
\right\rangle$ for all $S \in \mathcal{B}(\mathfrak{H}_s
\otimes\mathfrak{H}_s)$ for which
the summation converges. We observe that $D_e(I)  = \left\langle e \otimes
e, e \otimes e \right\rangle = 1$. If $d_e(I,Q)$ is
well defined for every projection $Q$ on $\mathfrak{H}_s
\otimes \mathfrak{H}_s$, then $D_e(Q)$ converges for every projection $Q$.

Let $U$ be the unitary on $\mathfrak{H}_s
\otimes \mathfrak{H}_s$ such that $U(e_n \otimes e_m) = e_m \otimes e_n$
for each $n$ and $m$. Then $U^2 = I$.  So $U$ is self adjoint. Let
$Q_U = \frac{1}{2}(U+I)$. Then $Q_U$ is a projection.
Assume now that $D_e(Q_U)$ is convergent. Then $D_e(2 Q_U -I)$
is convergent, i.e., \[
\sum_{j(2) = 1}^\infty \left\langle U(e_{j(2)} \otimes e),
e \otimes e_{j(2)} \right\rangle =
\sum_{j(2)=1}^\infty
\left\langle (e \otimes e_{j(2)}), e \otimes e_{j(2)}
\right\rangle = \sum_{j(2) = 1}^\infty 1 \]
is convergent. This is false. So $D_e(Q_U)$ is divergent. Thus
$d_e(I, Q_U)$ is not defined by the formula.
Thus we conclude that \textbf{the natural
formula for the homogeneous decoherence functional of standard
quantum mechanics does not induce a finitely valued decoherence
functional on the space of all histories} but rather a generalized
functional $\underline{d}_\rho : \mathcal{P} \left(
\mathcal{K}_{\: \: t_1, \cdots, t_n} \right) \times \mathcal{P} \left(
\mathcal{K}_{\: \: t_1, \cdots, t_n} \right) \to \mathbb{C} \cup \{
\infty \}$ with values in the Riemann sphere $\mathbb{C} \cup \{ \infty
\}$. \hfill $\Box$ \\ \\
The generalized decoherence functional $\underline{d}_\rho : \mathcal{P}
\left(
\mathcal{K}_{\: \: t_1, \cdots, t_n} \right) \times \mathcal{P} \left(
\mathcal{K}_{\: \: t_1, \cdots, t_n} \right) \to \mathbb{C} \cup \{
\infty \}$ of $d_\rho$ is given by
\[ \underline{d}_\rho(h,k) = \left\{ \begin{array}{r@{\quad:\quad}l}
d_\rho(h,k) & \mbox{whenever the series defining } d_\rho(h,k)
\mbox{ is well-defined and finite} \\ \infty & \mbox{else}
\end{array} \right. . \]
It is easy to see that our argument above already implies that
in standard quantum mechanics no standard decoherence functional
(of order two or greater)
over an infinite dimensional single-time Hilbert space
$\mathfrak{H}_s$ can be
extended to a completely additive finitely valued decoherence
functional on the space of all histories. For, by Wright \cite{Wright98},
such an extension would be bounded, contrary to the argument above.
In the next section it will become clear that even if the requirement
of complete additivity is dropped there is no hope of extending
$d_\rho$ to all histories. Our argument above also shows that there
are histories such that no decoherence functional $d_\rho$ assumes
a finite value on them. For instance, in our example above the
infinite sum defining $d_\rho(I,Q_U)$ diverges independently of $\rho$.
This result seems to indicate that the space $\mathcal{P} \left(
\mathcal{K}_{\: \: t_1, \cdots, t_n} \right)$ contains unphysical
elements and one might hope that the standard decoherence functional
is well defined on some suitably chosen smaller space of histories.

Bounded decoherence functionals have canonical representations
as quadratic forms on von Neumann algebras, see Corollary 4
\cite{Wright95}.
Surprisingly, in view of the negative results above, this is
almost true for standard decoherence functionals. It turns out
that each standard decoherence functional can be identified with
a positive, but unbounded, quadratic form defined on a `dense'
$^*$-subalgebra of $\mathcal{B}(\mathcal{H})$. This is clarified below.
\subsection{Representing standard decoherence functionals by unbounded
quadratic forms}
The non-boundedness of the standard decoherence functional forces
us to consider unbounded decoherence functionals as `necessary evils'
like unbounded operators. Also, like unbounded operators, their
domains are non-closed subspaces of the underlying Banach space.
In this subsection we shall show that in standard quantum mechanics
every decoherence functional $d_\rho$ associated with the initial state
$\rho$ can be represented by a unbounded quadratic form.

Specifically we consider the restriction of $d_\rho$ to histories of
order $n$ (which we denote also by $d_\rho$ - slightly
abusing the notation)
and call the resulting functional the decoherence functional
of order $n$. In this subsection we shall always assume that the
single time Hilbert space $\mathfrak{H}_s$
is infinite dimensional. We shall
see that for each such standard decoherence functional
$d_\rho$ of order $n$ on $\mathcal{H} = \mathfrak{H}_s \otimes
\cdots \otimes \mathfrak{H}_s$ there is a Hilbert space
\textsf{H} and
an operator $R_\rho : \mathcal{B}(\mathfrak{H}_s) \otimes_{alg}
\cdots \otimes_{alg}
\mathcal{B}(\mathfrak{H}_s) \to \mathsf{H}$ such that
\[ d_\rho(p_1 \otimes p_2 \otimes \cdots \otimes p_n,
q_1 \otimes q_2 \otimes \cdots \otimes q_n) = \left\langle R_\rho(p_1
\otimes p_2 \otimes \cdots \otimes p_n), R_\rho(q_1 \otimes q_2 \otimes
\cdots \otimes q_n) \right\rangle. \]
It follows that $d_\rho$ extends to a positive quadratic form
$D_\rho$ on $\mathcal{B}(\mathfrak{H}_s) \otimes_{alg} \cdots
\otimes_{alg} \mathcal{B}(\mathfrak{H}_s)$ where
\begin{equation} \label{E5}
D_\rho(v,w) = \left\langle R_\rho(v), R_\rho(w) \right\rangle,
\end{equation}
for each $v$ and each $w$ in the $n$-fold algebraic tensor product
of $\mathcal{B}(\mathfrak{H}_s)$ by itself.
We shall see below that when $\mathfrak{H}_s$ is infinite
dimensional, then $D_\rho$ is \emph{not} bounded
and that the map $R_\rho$ is an unbounded
operator. However, the representation (\ref{E5}) is a very close analogue
of the representations of bounded decoherence functionals obtained
in \cite{Wright95}. For, by the results of \cite{Wright95},
any bounded decoherence
functional on the projections of $H^\#$ (where $H^\#$ is a
Hilbert space with dimension greater
than 2) can be extended to a bounded quadratic form which can
be expressed as the difference of bounded semi-inner products
on $\mathcal{B}(H^\#)$. \\ \\
\noindent \emph{Remark}: For bounded
decoherence functionals, three notions of
positivity were distinguished in \cite{Wright98}. Only the strongest of
these corresponds to the representing quadratic form being positive.
However all decoherence functionals arising canonically in the
history formulation of standard quantum mechanics have this strong
positivity property although they fail to be bounded when the single
time Hilbert space $\mathfrak{H}_s$ is infinite dimensional. \\

Let $ \pi : \mathcal{B}(\mathfrak{H}_s)^n \to \mathcal{B}(\mathfrak{H}_s)$
be the product map defined by $\pi(x_1,x_2, \cdots, x_n)
= x_n x_{n-1} \cdots x_1$. Then, since $\pi$ is an $n$-linear
form, it follows by
the basic algebraic properties of tensor products, that there
exists a unique linear map $\Pi : \mathcal{B}(\mathfrak{H}_s) \otimes_{alg}
\cdots \otimes_{alg} \mathcal{B}(\mathfrak{H}_s) \to
\mathcal{B}(\mathfrak{H}_s)$ such that
\[ \Pi(x_1 \otimes x_2 \otimes \cdots \otimes x_n) =
\pi(x_1,x_2, \cdots, x_n) = x_n x_{n-1} \cdots x_1 \]
for all $x_1,x_2, \cdots, x_n \in \mathcal{B}(\mathfrak{H}_s)$.
Let $\phi$ be a normal state on $\mathcal{B}(\mathfrak{H}_s)$.
Then, for a unique positive trace class operator $\rho$ with trace
one, $\phi(x) = \mathrm{tr}_{\mathfrak{H}_s}(x \rho)$ for all $x$ in
$\mathcal{B}(\mathfrak{H}_s)$. The correspondence $\phi \leftrightarrow
\rho$ is a bijection. Then define $D_\phi$ on
$\mathcal{B}(\mathfrak{H}_s) \otimes_{alg} \cdots \otimes_{alg}
\mathcal{B}(\mathfrak{H}_s)$ by \[
D_\phi(z,w) = \phi(\Pi(w)^* \Pi(z)). \]
Let $p_1, p_2, \cdots, p_n$ and $q_1, q_2, \cdots, q_n$
be sequences of projections in
$\mathcal{B}(\mathfrak{H}_s)$. Then
\begin{eqnarray*}
D_\phi(p_1 \otimes p_2 \otimes \cdots \otimes p_n,
q_1 \otimes q_2 \otimes \cdots \otimes q_n) & = & \phi(q_1 q_2 \cdots q_n
p_n p_{n-1} \cdots p_1) \\ & = & \mathrm{tr}_{\mathfrak{H}_s}(q_1 q_2
\cdots q_n p_n p_{n-1} \cdots p_1 \rho). \end{eqnarray*}
But, by Proposition 5.2.2 of \cite{Pedersen79},
$\mathrm{tr}(ab \rho) = \mathrm{tr}(b \rho a)$, so
$D_\phi$ extends the standard decoherence functional $d_\rho$ arising from
$\rho$.

For any $z$ in $\mathcal{B}(\mathfrak{H}_s) \otimes_{alg} \cdots
\otimes_{alg} \mathcal{B}(\mathfrak{H}_s)$,
\[ D_\phi(z,z) = \phi(\Pi(z)^* \Pi(z)) \geq 0. \]
So $D_\phi$ is a semi-inner product on $\mathcal{B}(\mathfrak{H}_s)
\otimes_{alg} \cdots \otimes_{alg} \mathcal{B}(\mathfrak{H}_s)$.
Let $N_\phi = \{ z : \phi(\Pi(z)^* \Pi(z)) = 0 \}$.
It follows from the Cauchy-Schwarz
inequality that $N_\phi$ is a vector subspace of
$\mathcal{B}(\mathfrak{H}_s)
\otimes_{alg} \cdots \otimes_{alg} \mathcal{B}(\mathfrak{H}_s)$.
Let $R_\rho$ be the quotient map from
$\mathcal{B}(\mathfrak{H}_s)
\otimes_{alg} \cdots \otimes_{alg} \mathcal{B}(\mathfrak{H}_s)$
onto $(\mathcal{B}(\mathfrak{H}_s)
\otimes_{alg} \cdots \otimes_{alg} \mathcal{B}(\mathfrak{H}_s))/N_\phi$.
Then $D_\phi$ induces an inner product $\langle \cdot, \cdot \rangle$
on the quotient
such that $\left\langle R_\rho(z), R_\rho(w) \right\rangle = D_\phi(z,w)$.
Thus $(\mathcal{B}(\mathfrak{H}_s)
\otimes_{alg} \cdots \otimes_{alg} \mathcal{B}(\mathfrak{H}_s))/N_\phi$
is a pre-Hilbert space. Let \textsf{H} be the Hilbert space obtained by
completing this pre-Hilbert space. We have proved:
\begin{theo}
\label{T52} Given a standard decoherence functional $d_\rho$ of order
$n$ on $\mathcal{H} = \mathfrak{H}_s \otimes \mathfrak{H}_s \otimes
\cdots \otimes \mathfrak{H}_s$, there exists a Hilbert space
\emph{\textsf{H}} and a linear operator $R_\rho$ from
$\mathcal{B}(\mathfrak{H}_s) \otimes_{alg} \cdots
\otimes_{alg} \mathcal{B}(\mathfrak{H}_s)$
onto a dense subspace of \emph{\textsf{H}} such that \[
d_\rho(p_1 \otimes p_2 \otimes \cdots \otimes
\cdots \otimes p_n, q_1 \otimes
q_2 \otimes \cdots \otimes q_n) = \left\langle R_\rho( p_1 \otimes
p_2 \otimes \cdots \otimes p_n), R_\rho( q_1 \otimes q_2 \otimes \cdots
\otimes q_n) \right\rangle \]
for arbitrary projections $p_1, p_2, \cdots, p_n, q_1, q_2, \cdots, q_n$
in $\mathcal{B}(\mathfrak{H}_s)$.
Hence $d_\rho$ extends to a positive quadratic form $D_\rho$ on
$\mathcal{B}(\mathfrak{H}_s) \otimes_{alg} \cdots
\otimes_{alg} \mathcal{B}(\mathfrak{H}_s)$. \end{theo}
The following proposition shows that $D_\rho = D_\phi$ is unique.
\begin{prop}
\label{P53} Let $Q$ be a  sesquilinear form on $\mathcal{B}(\mathfrak{H}_s)
\otimes_{alg} \cdots \otimes_{alg} \mathcal{B}(\mathfrak{H}_s)$ such that
\[ Q(p_1 \otimes p_2 \otimes \cdots
\otimes p_n, q_1 \otimes q_2 \otimes
\cdots \otimes q_n) = d_\rho(p_1 \otimes p_2
\otimes \cdots \otimes p_n,
q_1 \otimes q_2 \otimes \cdots \otimes q_n) \]
for all projections $p_1, p_2, \cdots, p_n$ and $q_1, q_2, \cdots, q_n$
in $\mathcal{B}(\mathfrak{H}_s)$. Also let there exist a constant
$C$ such that \[ \left\vert
Q(x_1 \otimes x_2 \otimes \cdots \otimes x_n, y_1 \otimes y_2 \otimes
\cdots \otimes y_n) \right\vert \leq C \Vert x_1 \Vert
\Vert x_2 \Vert \cdots \Vert
x_n \Vert \Vert y_1 \Vert \Vert y_2 \Vert \cdots \Vert y_n \Vert.
\] Then $Q(u,v) = \left\langle R_\rho(u), R_\rho(v) \right\rangle =
D_\rho(u,v)$ for each $u$ and $v$
in $\mathcal{B}(\mathfrak{H}_s)
\otimes_{alg} \cdots \otimes_{alg} \mathcal{B}(\mathfrak{H}_s)$.
\end{prop}
\textbf{Proof}: Let us define a $2n$-linear form $L$ on
$\mathcal{B}(\mathfrak{H}_s)$ by \[
L(x_1, x_2, \cdots, x_n, y_1, y_2, \cdots, y_n) = Q(x_1 \otimes x_2
\otimes \cdots \otimes x_n, y_1^* \otimes y_2^* \otimes \cdots \otimes
y_n^*). \]
Then $L$ is bounded. Let $p_1, p_2, \cdots, p_n$ and $q_1, q_2,
\cdots, q_n$ be sequences of projections in $\mathcal{B}(\mathfrak{H}_s)$. Then
\[ L(p_1, p_2, \cdots, p_n, q_1, q_2, \cdots, q_n) =
\phi(q_1 q_2 \cdots q_n p_n p_{n-1} \cdots p_1). \]
It follows from the boundedness of $L$ and spectral theory that
\[ L(x_1, x_2, \cdots, x_n, y_1, y_2, \cdots, y_n) = \phi(y_1 y_2 \cdots
y_n x_n x_{n-1} \cdots x_1) \]
for all self-adjoint $x_1, x_2, \cdots, x_n, y_1, y_2, \cdots, y_n$.
It then follows
from multilinearity that this identity remains valid when $x_1, x_2,
\cdots, x_n, y_1, y_2, \cdots, y_n$ are arbitrary elements of
$\mathcal{B}(\mathfrak{H}_s)$. Thus
\begin{eqnarray*}
Q(x_1 \otimes x_2 \otimes \cdots \otimes x_n, y_1 \otimes y_2 \otimes
\cdots \otimes y_n) & = & L(x_1, x_2, \cdots, x_n, y_1^*, y_2^*,
\cdots, y_n^*) \\
& = & \phi( \pi(y_1, y_2, \cdots, y_n)^* \pi(x_1, x_2, \cdots, x_n))
\\ & = & \phi( \Pi(y_1 \otimes y_2 \otimes \cdots \otimes y_n)^*
\Pi(x_1 \otimes x_2 \otimes \cdots \otimes x_n)) \\
& = & \left\langle R_\rho(x_1 \otimes x_2 \otimes \cdots \otimes x_n),
R_\rho(y_1 \otimes y_2 \otimes \cdots \otimes y_n) \right\rangle
\end{eqnarray*}
Since each element of the algebraic tensor product
$\mathcal{B}(\mathfrak{H}_s)
\otimes_{alg} \cdots \otimes_{alg} \mathcal{B}(\mathfrak{H}_s)$
is a finite linear combination of simple tensors, $Q$
is of the required form. \hfill $\Box$ \\

The following proposition sheds further light on the unboundedness
results of Section \ref{S51}. In its statement we shall take $n = 2$
to simplify the notation but the result holds whenever $n \geq 2$.

Fix $\xi$, a unit vector in $\mathfrak{H}_s$. Let $\phi_\xi(x)
= \left\langle x \xi, \xi \right\rangle$ for each $x$ in
$\mathcal{B}(\mathfrak{H}_s)$. Let $D_\xi$ be constructed from $\phi_\xi$
as above.
\begin{prop}
\label{P54} Let $\mathfrak{H}_s$ be infinite dimensional. The positive
quadratic form $D_\xi$ is unbounded on $\mathcal{K}(\mathfrak{H}_s)
\otimes_{alg} \mathcal{K}(\mathfrak{H}_s)$. Hence $D_\xi$ is
unbounded on $\mathcal{B}(\mathfrak{H}_s) \otimes_{alg}
\mathcal{B}(\mathfrak{H}_s)$. \end{prop}
\textbf{Proof}: Let us assume that $D_\xi$ is bounded on
$\mathcal{K}(\mathfrak{H}_s)
\otimes_{alg} \mathcal{K}(\mathfrak{H}_s)$.
For each $z \in \mathcal{K}(\mathfrak{H}_s)
\otimes_{alg} \mathcal{K}(\mathfrak{H}_s)$
let $\delta(z) = D_\xi(z, 1)$.
Then $\delta$ is a bounded linear functional on
$\mathcal{K}(\mathfrak{H}_s)
\otimes_{alg} \mathcal{K}(\mathfrak{H}_s)$.
But $\delta(x \otimes y) = \phi_\xi(y x) = \langle x, y^* \rangle$.
So, by Proposition 0 \cite{Wright97} $\delta$ is
unbounded on $\mathcal{K}(\mathfrak{H}_s)
\otimes_{alg} \mathcal{K}(\mathfrak{H}_s)$.
This contradiction completes the proof. \hfill $\Box$ \\

The above proof is valid for standard decoherence functionals
corresponding to a vector state but, by a slight modification
of Proposition 0 \cite{Wright97}, a similar argument works for any standard
decoherence functional $d_\rho$ of order $n \geq 2$, provided
$\mathfrak{H}_s$ is infinite dimensional. \\ \\
\emph{Remark}: Proposition \ref{P54} implies that the quadratic form
$D_\xi$ is unbounded with respect
to any $C^*$-norm on $\mathcal{B}(\mathfrak{H}_s) \otimes_{alg}
\mathcal{B}(\mathfrak{H}_s)$. This is an immediate consequence
of the fact that, by nuclearity, all $C^*$-norms on
$\mathcal{B}(\mathfrak{H}_s) \otimes_{alg}
\mathcal{B}(\mathfrak{H}_s)$
coincide on $\mathcal{K}(\mathfrak{H}_s)
\otimes_{alg} \mathcal{K}(\mathfrak{H}_s)$.
\begin{co}
\label{C55} The map $\Pi : \mathcal{B}(\mathfrak{H}_s)
\otimes_{alg} \cdots \otimes_{alg} \mathcal{B}(\mathfrak{H}_s)
\to \mathcal{B}(\mathfrak{H}_s)$ is unbounded
if $\mathfrak{H}_s$ is infinite dimensional. The map $R_\rho :
\mathcal{B}(\mathfrak{H}_s)
\otimes_{alg} \cdots \otimes_{alg} \mathcal{B}(\mathfrak{H}_s)
\to \mathsf{H}$ is unbounded if $\mathfrak{H}_s$ is infinite dimensional.
\end{co}
\textbf{Proof}: If $\Pi$ or $R_\rho$ were bounded,
then $D_\rho$ would also be bounded. \hfill $\Box$
\begin{co}
\label{C56} Each standard decoherence functional $d_\rho$ of order
$n$, corresponding to a positive trace class operator $\rho$, has a
unique extension to a positive, quadratic form $D_\rho$ on the $n$-fold
algebraic tensor product $\mathcal{B}(\mathfrak{H}_s)
\otimes_{alg} \cdots \otimes_{alg} \mathcal{B}(\mathfrak{H}_s)$
such that \[ D_\rho(x,y) = \mathrm{tr}_{\mathfrak{H}_s}(\Pi(y)^* \Pi(x)
\rho). \] In particular $D_\rho(I,I) = 1$. \end{co}
\subsection{Existence of a generalized ILS-representation for the standard
homogeneous decoherence functional in infinite dimensions}
In Theorem \ref{T41} in Section IV we have seen that a general homogeneous
decoherence functional has an ILS-representation by a trace class
operator if and only if it is tensor bounded. However, we have
seen that the decoherence functional $d_\rho$
in standard quantum mechanics
associated with the initial state $\rho$ is not even bounded when the
single time Hilbert space $\mathfrak{H}_s$ is infinite dimensional.
Therefore
from the general results of Section IV.1 we cannot infer the
existence of an ILS-type representation for the standard decoherence
functional. However, the representation of the standard decoherence
functional given in Equation (\ref{E2}) is almost of the required form.
In the present subsection we prove the following theorem and
corollary (as always we denote the single time Hilbert space
by $\mathfrak{H}_s$)
\begin{theo}
\label{T57} Let $d_\rho^{hom}$ be the standard homogeneous
decoherence functional
of order $n$ in standard quantum mechanics associated with the
initial state $\rho$. There exists a unique bounded linear operator
$\mathfrak{M}_\rho$ on $\mathcal{H} \otimes \mathcal{H} =
\mathfrak{H}_s \otimes \cdots \otimes
\mathfrak{H}_s$ ($2n$ times) such that
\[ d_\rho^{hom}(p_1, \cdots, p_n, q_1, \cdots, q_n) =
\mathrm{tr}_{\mathcal{H} \otimes \mathcal{H}}\left( \left(p_1 \otimes
\cdots \otimes p_n \otimes q_1 \otimes \cdots \otimes q_n \right)
\mathfrak{M}_\rho \right) \]
whenever $p_i$ and $q_i$ are finite rank projections on
$\mathfrak{H}_s$ for all $i$. \end{theo}
Let us recall that Equation (\ref{E1})
implies that the standard decoherence
functional $d_\rho^{hom}$ is a homogeneous decoherence functional which is
completely additive in each of its $2n$ arguments. Thus we have
from Theorem \ref{T57}
\begin{co}
\label{C58} Let $d_\rho^{hom}$ be the standard homogeneous
decoherence functional
of order $n$ in standard quantum mechanics associated with the
initial state $\rho$. Then there exists a unique bounded linear operator
$\mathfrak{M}_\rho$ on $\mathcal{H} \otimes \mathcal{H} =
\mathfrak{H}_s \otimes \cdots \otimes
\mathfrak{H}_s$ ($2n$ times) such that, whenever
$P_i$ and $Q_i$ are projections in $\mathcal{P}(\mathfrak{H}_s)$ and
$(p_{i,r})$ and $(q_{i,r})$ are, respectively, orthogonal families
of finite rank projections with $P_i = \sum_r p_{i,r}$ and
$Q_i = \sum_r q_{i,r}$, then
\[ d_\rho^{hom}(P_1, \cdots, P_n, Q_1, \cdots, Q_n) =
\sum_{i_1, j_1} \cdots \sum_{i_n, j_n}
\mathrm{tr}_{\mathcal{H} \otimes \mathcal{H}}\left( \left(p_{1, i_1}
\otimes
\cdots \otimes p_{n, i_n} \otimes q_{1, j_1} \otimes \cdots \otimes
q_{n, j_n} \right)
\mathfrak{M}_\rho \right). \] \end{co}
\textbf{Proof}: The complete additivity of
$d_\rho^{hom}$ and Theorem \ref{T57} imply the
existence of a unique bounded linear operator $\mathfrak{M}_\rho$
with the required properties. \hfill $\Box$ \\ \\
\textbf{Proof} of Theorem \ref{T57}:
Let ${B}_\rho : \mathcal{B}(\mathfrak{H}_s)
\times \cdots \times \mathcal{B}(\mathfrak{H}_s) \to \mathbb{C}$
be the unique bounded $2n$-linear form which extends $d_\rho^{hom}$.
Let $\beta_\rho$ be the unique linear functional on
$\mathcal{K}(\mathfrak{H}_s)
\otimes_{alg} \cdots \otimes_{alg} \mathcal{K}(\mathfrak{H}_s)$
($2n$ times) such that \[
\beta_\rho(x_1 \otimes \cdots \otimes x_n \otimes y_1 \otimes
\cdots \otimes y_n) = B_\rho(x_1, \cdots, x_n, y_1, \cdots, y_n)
\] for all $x_i, y_i \in \mathcal{K}(\mathfrak{H}_s)$.

When $\xi$ is a unit vector in
$\mathfrak{H}_s \otimes_{alg} \cdots \otimes_{alg}
\mathfrak{H}_s$ ($2n$ times), then we denote by $p_\xi$ the
projection operator onto the subspace of
$\mathfrak{H}_s \otimes \cdots \otimes
\mathfrak{H}_s$  ($2n$ times) spanned
by $\xi$. The projection operator $p_\xi$ is in
$\mathcal{K}(\mathfrak{H}_s) \otimes_{alg} \cdots \otimes_{alg}
\mathcal{K}(\mathfrak{H}_s)$ ($2n$ times). Similarly, when $\xi_s$
is a unit vector in $\mathfrak{H}_s$, then we denote the
projection onto the subspace spanned by $\xi_s$ by $p_{\xi_s}$.

We shall see that for every positive trace
class operator $\rho$ with trace one the standard
decoherence functional $d_\rho$ is tracially bounded, i.e., there exists
a constant $C$ such that for every unit vector $\xi$ in
$\mathfrak{H}_s \otimes_{alg} \cdots \otimes_{alg}
\mathfrak{H}_s$ ($2n$ times) $\left\vert \beta_\rho \left( p_\xi \right)
\right\vert \leq C$. From Equation (\ref{E2}) we know that $\beta_\rho$
can be written as \[
\beta_\rho(P) = \sum_{j_1, \cdots, j_{2n} = 1}^\infty \omega_{j_1}
\left\langle P \left( \varepsilon_{j_1, \cdots, j_{2n}} \right),
\tilde{\varepsilon}_{j_1, \cdots, j_{2n}} \right\rangle. \]
For $\xi, \eta \in \mathfrak{H}_s \otimes \cdots \otimes \mathfrak{H}_s$
($2n$ times) let \[ S_\rho(\xi, \eta) := \sum_{j_1, \cdots, j_{2n} =
1}^\infty \omega_{j_1} \langle \varepsilon_{j_1, \cdots, j_{2n}},
\xi \rangle \langle \eta, \tilde{\varepsilon}_{j_1, \cdots,
j_{2n}} \rangle. \]
This expression is well defined since the sequences $\{ \langle
\varepsilon_{j_1, \cdots, j_{2n}}, \xi \rangle \}$ and $\{ \langle
\eta, \tilde{\varepsilon}_{j_1, \cdots, j_{2n}} \rangle \}$ are
square summable sequences in the Hilbert space $\ell_2(\mathbb{N}^{2n})$.
So, by the Cauchy-Schwarz inequality in $\ell_2(\mathbb{N}^{2n})$
\begin{eqnarray*}
\left\vert S_\rho(\xi, \eta) \right\vert & \leq & \sum_{j_1,
\cdots, j_{2n} =
1}^\infty \left\vert \langle \varepsilon_{j_1, \cdots, j_{2n}},
\xi \rangle \right\vert \left\vert
\langle \eta, \tilde{\varepsilon}_{j_1, \cdots,
j_{2n}} \rangle \right\vert \\
& \leq & \left(\sum_{j_1, \cdots, j_{2n} =
1}^\infty \left\vert \langle \varepsilon_{j_1, \cdots, j_{2n}},
\xi \rangle \right\vert^2 \right)^{\frac{1}{2}} \left(
\sum_{j_1, \cdots, j_{2n} = 1}^\infty \left\vert
\langle \eta, \tilde{\varepsilon}_{j_1, \cdots,
j_{2n}} \rangle \right\vert^2 \right)^{\frac{1}{2}} \\
& = & \Vert \xi \Vert  \Vert \eta \Vert
\end{eqnarray*}
So, $S_\rho$ is a bounded sesquilinear form on $\mathfrak{H}_s
\otimes \cdots \otimes \mathfrak{H}_s$ and thus there exists a
bounded linear operator $\mathfrak{M}_\rho$ on $\mathfrak{H}_s
\otimes \cdots \otimes \mathfrak{H}_s$ such that \[ S_\rho(\xi,
\eta) = \langle \mathfrak{M}_\rho \xi, \eta \rangle \] for all $\xi,
\eta$ in $\mathfrak{H}_s \otimes \cdots \otimes \mathfrak{H}_s$.
We note that from
$\Vert \mathfrak{M}_\rho \xi \Vert ^2 = \langle \mathfrak{M}_\rho
\xi, \mathfrak{M}_\rho \xi \rangle = \vert S_\rho(\xi,
\mathfrak{M}_\rho \xi) \vert \leq \Vert \xi \Vert \Vert
\mathfrak{M}_\rho \xi \Vert$ it follows that $\Vert \mathfrak{M}_\rho
\Vert \leq 1$.

For $\xi_1, \cdots, \xi_{2n} \in \mathfrak{H}_s$ let $\xi = \xi_1 \otimes
\cdots \otimes \xi_{2n}$. Then $p_\xi = p_{\xi_1} \otimes \cdots \otimes
p_{\xi_{2n}}$ and
\begin{eqnarray*}
\mathrm{tr}_{\mathcal{H} \otimes \mathcal{H}} \left(
(p_{\xi_1} \otimes \cdots \otimes p_{\xi_{2n}}) \mathfrak{M}_\rho
\right) & = & \langle \mathfrak{M}_\rho \xi, \xi \rangle \\
& = & \beta_{\rho}(p_{\xi_1} \otimes \cdots \otimes
p_{\xi_{2n}}) \\
& = & d_\rho \left( p_{\xi_1}, \cdots, p_{\xi_{n}}, p_{\xi_{n+1}}, \cdots,
p_{\xi_{2n}} \right) \end{eqnarray*}
Hence, by orthoadditivity,
$d_\rho(p_1, \cdots, p_n, q_1, \cdots, q_n) =
\mathrm{tr}_{\mathcal{H} \otimes \mathcal{H}}\left( \left(p_1 \otimes
\cdots \otimes p_n \otimes q_1 \otimes \cdots \otimes q_n \right)
\mathfrak{M}_\rho \right)$ \linebreak[4] whenever $p_i$
and $q_i$ are finite rank projections on $\mathfrak{H}_s$ for all
$i$. The uniqueness
of $\mathfrak{M}_\rho$ follows from the following lemma
\begin{lem}
\label{L59} Let $\mathcal{L}$ be a bounded operator on
$\mathfrak{H}_s \otimes \cdots \otimes
\mathfrak{H}_s$ such that, for all $\alpha_i \in \mathfrak{H}_s$,
\[ \left\langle \mathcal{L} (\alpha_1 \otimes \cdots \otimes
\alpha_n) , \alpha_1 \otimes \cdots \otimes \alpha_n \right\rangle
= 0. \] Then $\mathcal{L} = 0$. \end{lem}
Lemma \ref{L59} can be proved by iterating the argument given in the
proof of Lemma 4.1 in \cite{RudolphW97}. \\
This proves Theorem \ref{T57} and Corollary \ref{C58}. \hfill $\Box$
\subsection{Discussion}
We conclude this section with a discussion of the physical meaning
of histories with infinite weight. In the history reformulation of
standard quantum mechanics
the decoherence functional $d_\rho$ determines the consistent
sets of histories. In short, a subset $C$ of
$\mathcal{P} \left(\mathcal{K}_{\: \: t_1, \cdots, t_n} \right)$
is called consistent
if it is a Boolean lattice with respect to the lattice theoretical
operations induced from $\mathcal{P}
\left(\mathcal{K}_{\: \: t_1, \cdots, t_n}
\right)$ and if $\mathrm{Re } \: \: d_\rho(h,k) = 0$ for all disjoint
$h$ and $k$. In this case $p_\rho: C \to \mathbb{R}, p_\rho(h) :=
d_\rho(h,h)$ defines a probability
functional on $C$. Assertions about histories are meaningful only
with respect to a consistent set of histories and for
$h \in \mathcal{P} \left(\mathcal{K}_{\: \: t_1, \cdots, t_n} \right)$,
$d_\rho(h,h)$ can only be interpreted as probability of $h$
when explicit reference
is made to a fixed consistent set of histories.

Isham and Linden \cite{IshamL94}
have shown that already in the history formulation
of standard quantum mechanics over finite dimensional Hilbert
spaces the diagonal values of the standard decoherence functional
are greater than one for some inhomogeneous histories. In Section
\ref{S51} we have shown that for infinite dimensional Hilbert spaces
the standard decoherence functional is not even finitely valued
on the space of all histories. Clearly, a value $d_\rho(h,h) > 1$
cannot be interpreted as a probability for the inhomogeneous
history $h$.

We propose the following physical interpretation of inhomogeneous
histories with $d_\rho(h,h) > 1$: If $d_\rho(h,h) > 1$,
then $h$ is a coarse-graining
of mutually  exclusive histories, whose ``space-time'' interference
(measured by the decoherence functional $d_\rho$) is so large that
they cannot be distinguished as separate ``events'' in space-time.

The physical point at stake is that histories, which are disjoint
(i.e., which are represented by orthogonal projections) may nevertheless
have a large 'overlap' (since histories are spread out in time
two homogeneous histories can represent exclusive propositions
at some time and non-exclusive propositions at another time).
Accordingly for a pair of disjoint histories $h$ and $k$ which have
a large overlap in this sense, the coarse graining $h \vee k$
(representing
the proposition that the history $h$ or the history $k$ is realized)
may not represent a physically sensible proposition.

As a consequence of this, all histories with $d_\rho(h,h) > 1$ represent
unphysical propositions in the state  and must be dismissed.
The same is true for histories $h$ for which $d_\rho(h,h)$ is infinite.
Such histories represent no greater conceptual problem in this
interpretation than histories with $1 < d_\rho(h,h) < \infty$.

The axioms characterising a history quantum theory over an orthoalgebra
or over an effect algebra are abstracted from the history reformulation
of standard quantum mechanics over some Hilbert space. These
axioms were first given by Isham \cite{Isham94}. However, in the past it
has always been assumed that a decoherence functional is a complex
valued function on pairs of histories. This choice can be motivated
by appealing to the history reformulation of standard quantum
mechanics over finite-dimensional Hilbert spaces. However, our
analysis above of the history reformulation of standard quantum
mechanics over infinite dimensional Hilbert spaces has shown
that the standard decoherence functional is unbounded and that
its extension to the space of all histories would in general
take values in the Riemann sphere. Accordingly, one has to expect
that also in general history quantum theories the decoherence
functional will in general be a function with values in the Riemann
sphere (or, equivalently, represented by an unbounded `densely'
defined quadratic form).

\subsubsection*{Acknowledgements}
Oliver Rudolph is a Marie Curie Research Fellow
and carries out his research at the Imperial College as part
of a European Union training project financed by the European
Commission under the Training and Mobility of Researchers (TMR)
programme.

\end{document}